\documentclass[%
 aip,
 amsmath,amssymb,
 reprint,%
]{revtex4-1}

\usepackage{graphicx}
\usepackage{dcolumn}
\usepackage{bm}

\usepackage{amsmath}
\usepackage{graphicx}
\usepackage{ dsfont }
\usepackage{dcolumn}
\usepackage{bm}
\usepackage{color,soul}
\usepackage{xcolor}
\usepackage{url}
\usepackage{tabularx}
\usepackage{array}
\usepackage{booktabs}
\usepackage{comment}
\usepackage{hyperref}
\usepackage{footnote}
\usepackage{lipsum}
\usepackage[normalem]{ulem}

\usepackage[utf8]{inputenc}
\usepackage[T1]{fontenc}
\usepackage{mathptmx}
\usepackage{etoolbox}

\newcounter{firstbib}

\makeatletter
\def\@email#1#2{%
 \endgroup
 \patchcmd{\titleblock@produce}
  {\frontmatter@RRAPformat}
  {\frontmatter@RRAPformat{\produce@RRAP{*#1\href{mailto:#2}{#2}}}\frontmatter@RRAPformat}
  {}{}
}%
\makeatother
\begin{document}

\preprint{AIP/123-QED}

\title[Diversity of states in a chiral magnet nanocylinder]{Diversity of states in a chiral magnet nanocylinder}

\author{Andrii~S.~Savchenko}
\email{a.savchenko@fz-juelich.de} 
 \affiliation{Peter Gr\"unberg Institute and Institute for Advanced Simulation, Forschungszentrum J\"ulich and JARA, 52425 J\"ulich, Germany}
 \affiliation{Donetsk Institute for Physics and Engineering, National Academy of Sciences of Ukraine, 03028 Kyiv, Ukraine}

\author{Fengshan~Zheng}
\email{f.zheng@fz-juelich.de}
\affiliation{Ernst Ruska-Centre for Microscopy and Spectroscopy with Electrons and Peter Gr\"unberg Institute, Forschungszentrum J\"ulich, 52425 J\"ulich, Germany}
\affiliation{Spin-X Institute, School of Physics and Optoelectronics, State Key Laboratory of Luminescent Materials and Devices, Guangdong-Hong Kong-Macao Joint Laboratory of Optoelectronic and Magnetic Functional Materials, South China University of Technology, Guangzhou 511442, China}

\author{Nikolai~S.~Kiselev}
\email{n.kiselev@fz-juelich.de}
\affiliation{Peter Gr\"unberg Institute and Institute for Advanced Simulation, Forschungszentrum J\"ulich and JARA, 52425 J\"ulich, Germany}

\author{Luyan~Yang}
\affiliation{Ernst Ruska-Centre for Microscopy and Spectroscopy with Electrons and Peter Gr\"unberg Institute, Forschungszentrum J\"ulich, 52425 J\"ulich, Germany}

\author{Filipp~N.~Rybakov}
\affiliation{Department of Physics and Astronomy, Uppsala University, SE-75120 Uppsala, Sweden}

\author{Stefan~Bl\"ugel}
\affiliation{Peter Gr\"unberg Institute and Institute for Advanced Simulation, Forschungszentrum J\"ulich and JARA, 52425 J\"ulich, Germany}

\author{Rafal~E.~Dunin-Borkowski}
\affiliation{Ernst Ruska-Centre for Microscopy and Spectroscopy with Electrons and Peter Gr\"unberg Institute, Forschungszentrum J\"ulich, 52425 J\"ulich, Germany}

\date{\today}

\begin{abstract}
The diversity of three-dimensional magnetic states in an FeGe nanocylinder is studied using micromagnetic simulations and off-axis electron holography in the transmission electron microscope.
In particular, we report the observation of a dipole string -- a spin texture composed of two coupled Bloch points -- which becomes stable under geometrical confinement.
Quantitative agreement is obtained between experimental and theoretical phase shift images by taking into account the presence of a damaged layer on the surface of the nanocylinder.
The theoretical model is based on the assumption that the damaged surface layer, which results from focused ion beam milling during sample preparation, has similar magnetic properties to those of an amorphous FeGe alloy.  
The results highlight the importance of considering the magnetic properties of the surface layers of such nanoscale samples, which influence their magnetic states.
\end{abstract}

\maketitle

\section{\label{sec:level1}INTRODUCTION}

\begin{figure}
\includegraphics[width=8cm]{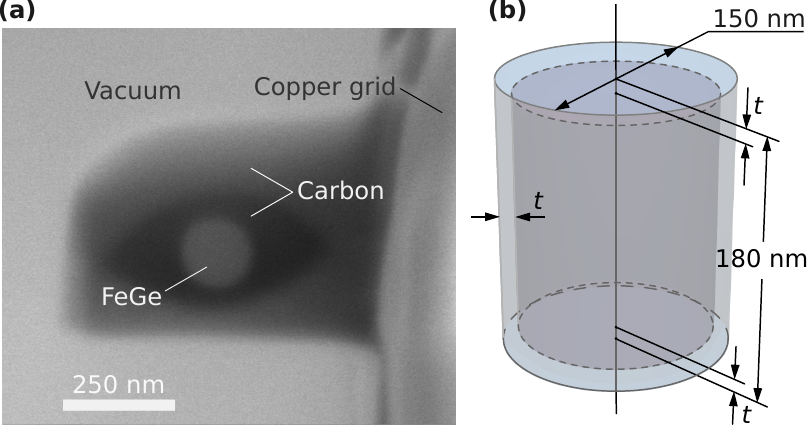}
\caption{ 
(a) Scanning electron micrograph of an FeGe nanocylinder cut from a bulk crystal using FIB milling. (See the Supplemental video in Ref.~\cite{Zheng_17}). 
The image was recorded using a short exposure time to avoid damage to the sample by the electron beam.
C~coating was used to minimize damage to the sample during FIB preparation.
 (b)~Schematic illustration of the sample geometry. $t$ is the thickness of the damaged layer formed due to FIB sample preparation.
\label{Fig_1}
}
\end{figure}

Cubic chiral magnets with bulk Dzyaloshinskii-Moriya interaction (DMI)~\cite{Dzyal_58, Moriya_60} are known to host a wide range of noncollinear magnetic textures, such as helical spirals~\cite{Beille_83, Uchida_06}, skyrmions~\cite{Bogdanov_89,Yu_10,Du_18_int}, antiskyrmions~\cite{F_Zheng_21}, skyrmion bags~\cite{Rybakov_19,Foster_19,Tang_21}, braids~\cite{Sk_braids}, chiral bobbers~\cite{Rybakov_2015, Zheng_18} and heliknotons~\cite{Voinescu_20}.
The most commonly used method to visualize such magnetic states is transmission electron microscopy (TEM), which is able to provide nanometer-size spatial resolution in real space.
Whereas conventional Lorentz TEM has been widely used for observations of magnetic states in extended plates~\cite{Yu_10, Du_18_int}, it is less suitable for studies of geometrically-confined nanoscale samples, such as disks~\cite{Zheng_17} and stripes~\cite{Jin_17}, in part because the \textit{out-of-focus} imaging condition results in the presence of Fresnel fringes at the sample edges, which can dominate the magnetic contrast of primary interest.
Contrast arising at the sample edges can be subtracted more easily by using the \textit{in-focus} technique of off-axis electron holography, which provides direct access to all spatial frequencies of the electron optical phase shift~\cite{Lichte_08,Midgley_09}.
Here, we use off-axis electron holography technique to study an FeGe nanocylinder that has a diameter of $150$~nm, a thickness of $180$~nm and is prepared using focused ion beam (FIB) milling, as shown in Fig.~\ref{Fig_1}~(a).
In contrast to prior work on an FeGe nanodisk~\cite{Zheng_17} that had a diameter of $160$~nm and a thickness of $90$~nm, different magnetic states are observed in this cylindrical sample.

Following a well-established approach based on micromagnetic simulations, we calculate theoretical phase shift images corresponding to the lowest energy states and compare them to experimental phase images recorded using off-axis electron holography.
Our results show that magnetic states in nanoscale samples cut from bulk crystals using FIB milling can be strongly affected by the presence of a damaged surface layer, as schematically shown in Fig.~\ref{Fig_1}(b).
We assume that (i)~the damaged layer has a disordered or amorphous structure, (ii)~due to the absence of crystal order in the damaged layer, the DMI is negligibly small and (iii)~the net magnetization and Heisenberg exchange coupling are nearly the same as those in the inner, undamaged region of B20-type crystal.
By varying the thickness of the damaged layer in our micromagnetic model, we obtain quantitative agreement between theoretical phase shift images and experimental measurements.
One of the experimentally-observed states is identified as an exotic magnetic state composed of two Bloch points of opposite topological index, which is known in the literature as a toron~\cite{Leonov_18, Smalyukh_10}, monopole-antimonopole pair~\cite{Liu_18} and dipole string~\cite{Ostlund_81, Calderon_01, Muller_20}.

\section{Micromagnetic calculations}

For theoretical analysis of magnetic states in a sample of confined geometry, we use a standard model for isotropic chiral magnets that includes Heisenberg exchange, DMI and demagnetizing fields.
The energy density functional can be written in the form~\cite{Sk_braids,Fratta}
\begin{align} 
\mathcal{E}\!=\!\int\limits_{V_\mathrm{m}}\!d\mathbf{r}\ 
&\mathcal{A}\sum\limits_{i=x,y,z} |\nabla m_i|^2 
+\mathcal{D}\,\mathbf{m}\!\cdot(\nabla\!\times\!\mathbf{m})
- M_\mathrm{s}\,\mathbf{m}\!\cdot\!\mathbf{B} + \nonumber\\
&+ \frac{1}{2\mu_0}\int\limits_{\mathbb{R}^3}\!d\mathbf{r}\ 
\sum\limits_{i=x,y,z}|\nabla A_{\mathrm{d}, i}|^2~,
\label{Ham_m}
\end{align}
where ${\mathbf{m}(\mathbf{r})}=\mathbf{M}(\mathbf{r})/M_\text{s}$ is the magnetization unit vector field, $M_\text{s}$ is the saturation magnetization, $\mu_0$ is the vacuum permeability,
${\mathbf{A}_\text{d}(\mathbf{r})}$ is the component of magnetic vector potential induced by the magnetization and $\mathcal{A}$ and $\mathcal{D}$ are the Heisenberg exchange constant and DMI constant, respectively.
In general, it is assumed that the magnetic field $\mathbf{B}(\mathbf{r})$ is the sum of the external magnetic field $\mathbf{B}_\mathrm{ext}$ and the demagnetizing field produced by the sample itself, \emph{i.e.},  $\mathbf{B}=\mathbf{B}_\text{ext}+\nabla\!\times\!\mathbf{A}_\text{d}$.

Here, we only consider the case of zero external magnetic field, $B_\text{ext}=0$.
The last two terms in Eq.~\eqref{Ham_m} therefore represent the demagnetizing field energy. 
We used the following material parameters for FeGe~\cite{Zheng_18, Sk_braids, F_Zheng_21}:
$\mathcal{A}~=~4.75$~pJm$^{-1}$, $\mathcal{D}~=~0.853$~mJm$^{-2}$ and $M_\text{s}~=~384$~kAm$^{-1}$.
The sample is assumed to have a cylindrical shape with a diameter of 150~nm and a height of 180~nm, as shown schematically in Fig.~\ref{Fig_1}(b).

The damaged surface layer of thickness $t$ is assumed to have the same material parameters as above, but with the DMI set to zero, which can be justified as follows.
Because of the damage produced by Ga ions during FIB preparation, the ordered crystal structure that is present in the sub-surface layers of the sample is assumed to be broken.
The properties of such a disordered state are expected to be similar to those of an amorphous FeGe alloy, which are well studied.
In particular, it is known that, at nearly equal concentrations of Fe and Ge, the Curie temperature of amorphous FeGe~\cite{Choe_96, Suran_76} is close to that of the corresponding crystalline B20 structure~\cite{Lundgren_68}.
At the same time, amorphous FeGe does not exhibit non-collinear magnetic order, at least at temperatures below $T~=~130$~K~\cite{Streubel_21}.
It is therefore reasonable to expect that the exchange stiffness in amorphous FeGe is close to that in a B20-type crystal, but that the contribution of the DMI is negligibly small.
The saturation magnetization of amorphous FeGe at
$T~=~100$~K\cite{Suran_76, Streubel_21} is also reported to be nearly identical to that of B20-type FeGe.

The mesh size in our simulations was fixed to be 1~nm in all spatial directions.
This choice allows the impact of the damaged layer to be studied by varying its thickness in steps of 1~nm.
Stable solutions of the functional \eqref{Ham_m} were found by numerical energy minimization using the Excalibur code~\cite{Excalibur}. 
In order to ensure reproducibility of the results, we also performed a check using publicly-available Mumax~\cite{mumax} code. The corresponding Mumax script is provided in the Supplementary Material.

\begin{figure*}
\includegraphics[width=18cm]{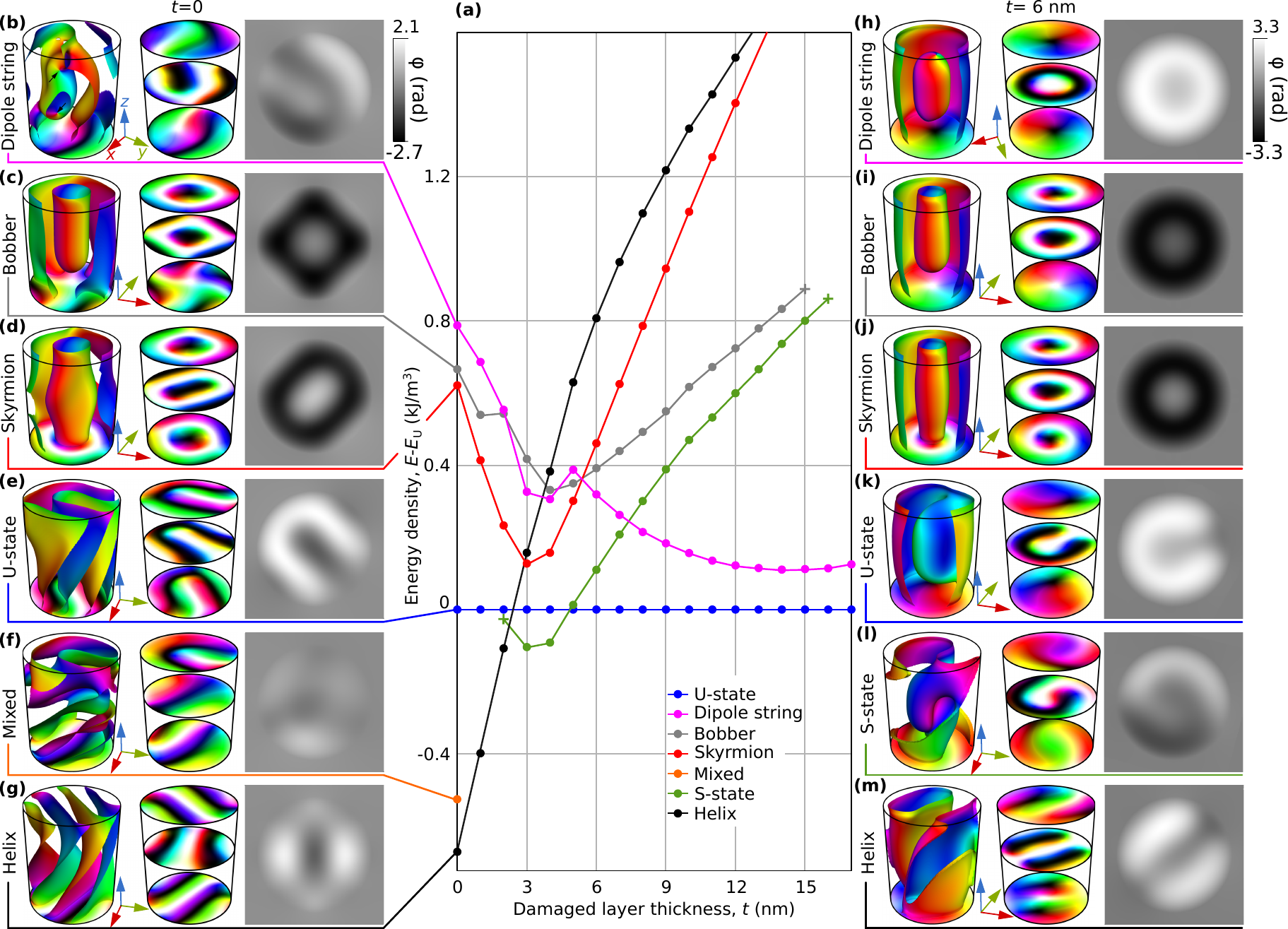}
\caption{ 
Results of micromagnetic simulations for the FeGe nanocylinder for different thicknesses of the damaged surface layer. 
(a)~Energy densities of representative stable magnetic configurations plotted as a function of damaged layer thickness with respect to the energy of the U-state shown in~(e).
Crosses indicate the critical damaged layer thickness above (or below) which the state becomes unstable.
The left and center images in (b)-(m) show the magnetization vector field as isosurfaces $m_z=0$ and cross-sections of the simulated cylindrical domain in three equidistant planes, respectively.
The black arrows in (b) mark the positions of Bloch points.
The right images in (b)-(m) shows the calculated electron phase shift $\varphi$ when the electron beam is parallel to the main axis of the cylinder.
The gray scale bar in (b) applies to all of the images in (b)-(g). The gray scale bar in (h) applies to all of the images in (h)-(m).
\label{Fig_2}
}
\end{figure*}

In order to identify the lowest energy states in the FeGe nanocylinder, we performed a large number of energy minimization cycles using different initial states.
Representative results of these calculations are shown in Fig.~\ref{Fig_2} for states with the lowest energies.
The magnetic texture of each state in Fig.~\ref{Fig_2}~(b)-(m) is shown in the form of both an isosurface $m_{z}=0$ and a cross-section of the simulated domain in the top, middle and bottom planes. 
The color code reflects the direction of the magnetization according to the standard color scheme used in Mumax, where white and black denote $m_{z}=1$ and $m_{z}=-1$, respectively, while red-green-blue correspond to the angle between the $m_{xy}$ projection of the magnetization and the $x$~axis.
The third image in each row is the electron optical phase shift $\varphi$ calculated with the incident electron beam antiparallel to the $z$~axis, according to the expression~\cite{Graef_2001}
\begin{align} 
\varphi (x,y)\!=\!\frac{2\pi e}{h}\int\limits_{-\infty}^{+\infty}\!dz\ 
\mathbf{A}_\text{d}(\mathbf{r})\cdot \hat{\mathbf{e}}_\text{z}~,
\label{Ph_shift}
\end{align}
where $e$ is a positive elementary charge and $h$ is Planck's constant.
The results shown in (b)-(g) correspond to the case of no damaged layer (\emph{i.e.}, $t~=~0$), while the results shown in (h)-(m) correspond to a damaged layer of thickness $t~=~6$~nm. 

Figure~\ref{Fig_2}(a) summarises the energies of the different states calculated as a function of damaged layer thickness.
The energies are provided as differences with respect to the energy of the U-state, which represents the ground state of the system in the case of a thick damaged layer.
The diagram shows that the energies depend strongly on the thickness of the damaged layer.
States that contain Bloch points, the dipole string depicted in (b) and (h) or the chiral bobber depicted in (c) and (i) never appear to be lowest energy states, as expected since such singularities typically lead to an increase in Heisenberg exchange, which is the leading energy term. 
In contrast, for $t\geq8$~nm the dipole string becomes a metastable state, whose energy is slightly higher than that of the ground state and much lower than the energies of other states (\emph{e.g.}, a skyrmion).
The energy of the dipole string reaches its minimum at $t\approx14$~nm and then increases with $t$.
Remarkably, in contrast to the situation for a 90-nm-thick FeGe nanodisk~\cite{Zheng_17}, the skyrmion is never the ground state of the system, either in the presence or in the absence of a damaged layer.
For $t\geq6$~nm, the energy of a skyrmion is even higher than that of a chiral bobber. 
The U-state in (e) and (k) can be considered as the helical state depicted in (g) and (m), but slightly distorted by additional modulations across the thickness.
The mixed state in (f) corresponds to a mixture of a helix and a U-state and has an energy in between these two states.
In the range $0<t<3$ nm, a few states correspond to mixtures of a U-state and a helix.
However, they are not shown in the figure, as their energies are always higher than the helix energy. 

The magnetic configuration corresponding to the S-state shown in Fig.~\ref{Fig_2}(l) can, in principle, also be thought of as a distorted helix.
A distinctive feature of this state is the characteristic contrast of the phase shift, which resembles a Yin-Yang symbol.
In the next section, we show that such contrast is observed experimentally using off-axis electron holography.
Another remarkable feature of the S-state is its finite range of existence.
It remains stable only in the $3$ nm $\leq t\leq 16$~nm range of damaged layer thickness.
For $t<3$~nm, it converges to a helix or one of the mixed states, while for $t>16$~nm it converges to a U-state.

\begin{figure*}
\includegraphics{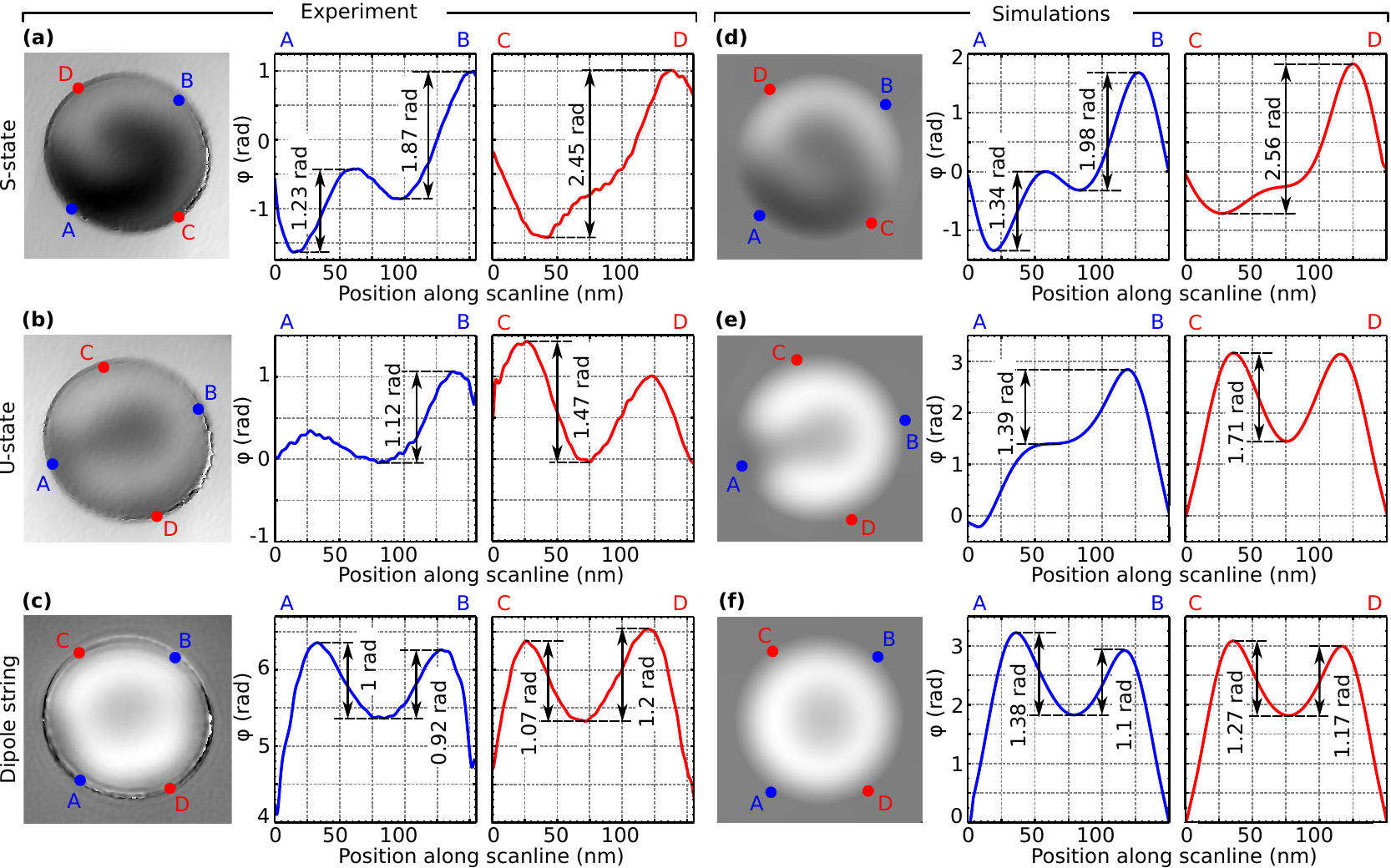}
\caption{ 
Quantitative comparisons between experimental and theoretical phase shift images for the FeGe nanocylinder. The left column in (a)-(c) shows experimental images. The left column in (d)-(f) shows corresponding theoretical images calculated for a damaged layer thickness $t=6$~nm. The latter images are identical to those shown in Fig.~\ref{Fig_2}(l), (k), and (h), respectively. Corresponding phase shift profiles along the lines $AB$ and $CD$ are shown in the center and right columns.
\label{Fig2_Holog}
}
\end{figure*}

Since we consider only the case of $B_\mathrm{ext}=0$, each of the states shown in Fig.~\ref{Fig_2} has an energetically-equivalent state with opposite net magnetization.
For instance, the magnetization in the core of the dipole string depicted in (h) is along the positive direction of the $z$-axis, but the dipole string with opposite magnetization in the center has the same energy.
Since the DMI conserves the chirality of the magnetization, the phase shift images of such configurations are distinct only by an inversion of the contrast. 

In the next section, we compare experimental measurements with the micromagnetic simulations.

\section{Experimental observations}

Magnetic states were recorded using off-axis electron holography in an FEI Titan G$^2$ 60-300 TEM operated at 300~kV. The microscope was operated in Lorentz mode with the objective lens switched off to provide magnetic-field-free conditions at the position of the specimen.
A liquid N$_2$ specimen holder (Gatan model 636) was used to control the sample temperature to be approximately 95~K.
Off-axis electron holograms were recorded using an electrostatic biprism positioned in the conventional selected-area aperture plane of the microscope on a 4k~$\times$~4k Gatan K2 IS direct electron counting detector. 30 object holograms and 30 vacuum reference holograms, with a 4~s exposure time for each hologram, were recorded for each magnetic state to improve the signal-to-noise ratio. 
The holograms were analysed using a standard fast Fourier transform algorithm. In order to remove the mean inner potential (MIP) contribution to the total phase shift, the phase images were recorded both at low temperature and at room temperature. They were then aligned and subtracted from each other on the assumption that the MIP contribution was the same and that there were no significant changes in specimen charging or dynamical diffraction.

Examples of experimentally-measured phase images are shown in Figs~\ref{Fig2_Holog}(a)-(c), alongside phase line profiles extracted along segments $AB$ and $CD$ passing through the minimum and maximum values of $\varphi$.
Different states were observed on performing multiple cycles of demagnetization of the sample.
Two of these states, which are shown in Figs~\ref{Fig2_Holog}(a) and (b), appeared with nearly identical frequency, while the state shown in Fig.~\ref{Fig2_Holog}(c) were observed more seldom.
We infer that the magnetic states in (a) and (b) are those with the lowest energies, while that in (c) is metastable.
Whereas the energies of the states in (a) and (b) are thought to be comparable, it is difficult to identify which of them represents the ground state.

The theoretical images for the state calculated without a damaged layer shown in Figs~\ref{Fig_2}(b)-(g) do not fit the experimental images well. (See also Supplementary Fig.~S1). 
There are some similarities between Fig.~\ref{Fig_2}(b) and the experimental image shown in Fig.~\ref{Fig2_Holog}(a).
However, the dipole string without a damaged layer ($t~=~0$) represents a high energy state and thus should not appear experimentally more often than lower energy states.
Moreover, a quantitative comparison of the phase shift signal in Fig.~\ref{Fig_2}(b) does not match the experimental images. (See Supplementary Fig.~S1 for details).
In contrast, the micromagnetic simulations calculations for a damaged layer thickness of 6~nm show good agreement with the experimental results.
In particular, the experimental phase image and corresponding line profiles in (a) match the calculations for the S-state in (d).
It should be noted that only phase shift differences across each image, rather than the absolute value of the phase shift, are relevant for such comparisons.
The experimental results in (b), in turn, perfectly fit the U-state shown in (e).
Finally, the theoretical image for the dipole string in (f) provides a best fit to the experimental image shown in (c).
Although the theoretical images of the chiral bobber and skyrmion shown in Figs~\ref{Fig_2}(h) and (i) also provide contrast similar to that in Fig.~\ref{Fig2_Holog}(c), only the dipole string fits the experimental data on a quantitative level. (See Supplementary Fig.~S2).
Moreover, according to the micromagnetic simulations the energies of the skyrmion and chiral bobber for $t~=~6$~nm are higher than that of dipole string.
Therefore, it is reasonable to expect that the dipole string occurs with a higher probability than the skyrmion or chiral bobber.

In our experimental measurements, we also imaged a state whose contrast and absolute values of phase shift were close to those expected for a chiral bobber. (See Supplementary Fig.~S3). Because of elliptical distortions present in this experimental image, we could not state with confidence that this phase image corresponds to a chiral bobber. We therefore do not draw this conclusion in the present work.
In agreement with the theoretical prediction, we did not experimentally observe a skyrmion state, whose energy is even higher than that of the dipole string and the chiral bobber.  

Small discrepancies between the measured and calculated phase shift differences can be attributed to imperfections of the sample geometry and local deviations in the thickness of the damaged layer in the real sample~\cite{Giannuzzi_99, Du_18}.
In addition, material parameters such as $M_s$ and $\mathcal{A}$ in the damaged layer may not be identical to those for B20-type FeGe.
The damaged layer may contain $\text{Ga}^{+}$ ions and oxygen.
Its stoichiometric composition may also deviate slightly from that of B20-type FeGe~\cite{Wolf_21}.
However, we can exclude the possibility that the damaged layer is nonmagnetic. Details of corresponding calculations are provided in Supplementary Fig.~S4.
As the thickness of such a FIB-damaged layer is typically 4-10~nm \cite{Du_18,Wolf_21}, our estimation of $t~=~6$~nm fits this range.

\section{Conclusions}

Experimental electron optical phase shift images of a nanocylinder of B20-type FeGe recorded using off-axis electron holography have been interpreted using a micromagnetic model that takes into account the presence of a surface damaged layer resulting from the use of FIB milling during sample preparation.
On the assumption that the damaged layer is disordered and effectively behaves as amorphous FeGe, we obtain quantitative agreement between experimental and theoretical images for an estimated thickness of the damaged layer of $\sim6$~nm.

One of the experimentally observed states is identified as a dipole string (or toron) -- a spin configuration that is composed of two Bloch points of opposite topological charge.
To the best of our knowledge, this is the first direct observation of such a state.
An alternative approach for stabilizing a dipole string state was presented by Liu et al.~\cite{Liu_18}, who suggested that it can be stabilized in a nanocylinder of a chiral magnet by perpendicular pinning of the magnetization at the top and bottom bases of the cylinder.
Such pinning was assumed to be introduced by an  additional magnetic layer with strong perpendicular magnetic anisotropy sputtered onto the bases of the nanocylinder.
Here, we show instead that a dipole string can be stabilized naturally by effectively softening the surface layer by introducing damage by FIB milling.
Moreover, we show that even a small variation in the damage layer thickness can change the energy balance across a large diversity of possible magnetic states.
Further studies of the properties of such damaged layers may suggest practical applications for the reliable control of different magnetic states in magnetic nanostructures.

\section{Supplementary Material}
See supplementary material for Figs.~S1-S4  and Mumax script.

\section{Acknowledgments}
The authors acknowledge financial support from the European Research Council (ERC) under the European Union’s Horizon 2020 research and innovation program (Grant No. 856538, project “3D MAGiC”) and the Deutsche Forschungsgemeinschaft (DFG, German Research Foundation) – Project-ID 405553726 – TRR 270.
F.N.R. acknowledges support from the Swedish Research Council.


\end{document}



\title[]{Supplemental Material for ``Diversity of states in chiral magnet nanocylinder''}

\author{Andrii~S.~Savchenko}
 \affiliation{Peter Gr\"unberg Institute and Institute for Advanced Simulation, Forschungszentrum J\"ulich and JARA, 52425 J\"ulich, Germany}
 \affiliation{Donetsk Institute for Physics and Engineering, National Academy of Sciences of Ukraine, 03028 Kyiv, Ukraine}

\author{Fengshan~Zheng}
\affiliation{Ernst Ruska-Centre for Microscopy and Spectroscopy with Electrons and Peter Gr\"unberg Institute, Forschungszentrum J\"ulich, 52425 J\"ulich, Germany}

\author{Nikolai~S.~Kiselev}
\affiliation{Peter Gr\"unberg Institute and Institute for Advanced Simulation, Forschungszentrum J\"ulich and JARA, 52425 J\"ulich, Germany}

\author{Luyan~Yang}
\affiliation{Ernst Ruska-Centre for Microscopy and Spectroscopy with Electrons and Peter Gr\"unberg Institute, Forschungszentrum J\"ulich, 52425 J\"ulich, Germany}

\author{Filipp~N.~Rybakov}
\affiliation{Department of Physics and Astronomy, Uppsala University, SE-75120 Uppsala, Sweden}

\author{Stefan~Bl\"ugel}
\affiliation{Peter Gr\"unberg Institute and Institute for Advanced Simulation, Forschungszentrum J\"ulich and JARA, 52425 J\"ulich, Germany}

\author{Rafal~E.~Dunin-Borkowski}
\affiliation{Ernst Ruska-Centre for Microscopy and Spectroscopy with Electrons and Peter Gr\"unberg Institute, Forschungszentrum J\"ulich, 52425 J\"ulich, Germany}



\maketitle

\onecolumn

\begin{figure*}
\includegraphics{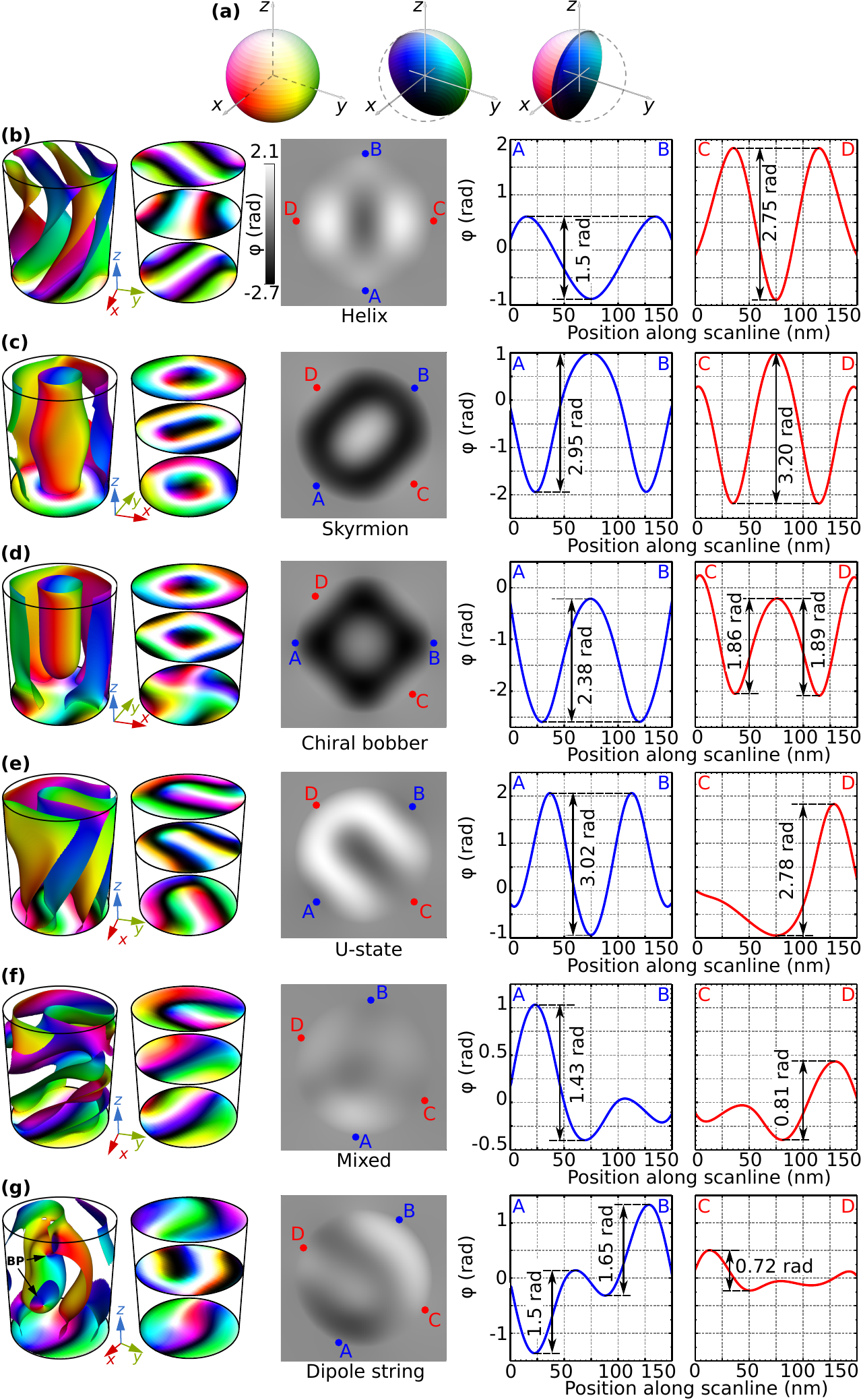}
\caption{\label{FigS1}
(a) Color code for the unit magnetization vector field. (b)-(g) Equilibrium magnetic states and corresponding electron phase images for the FeGe nanocylinder of 150 nm diameter and 180 nm height. In the first column we show the isosurfaces for $m_z=0$, in the second column we show the magnetization crossesctions in the top, bottom and middle planes. The calculated electron phase images and corresponding values of the magnetic phase shift along the line segments $AB$ and $CD$ are in the third, fourth and fifth columns.}
\end{figure*}

\begin{figure*}
\includegraphics{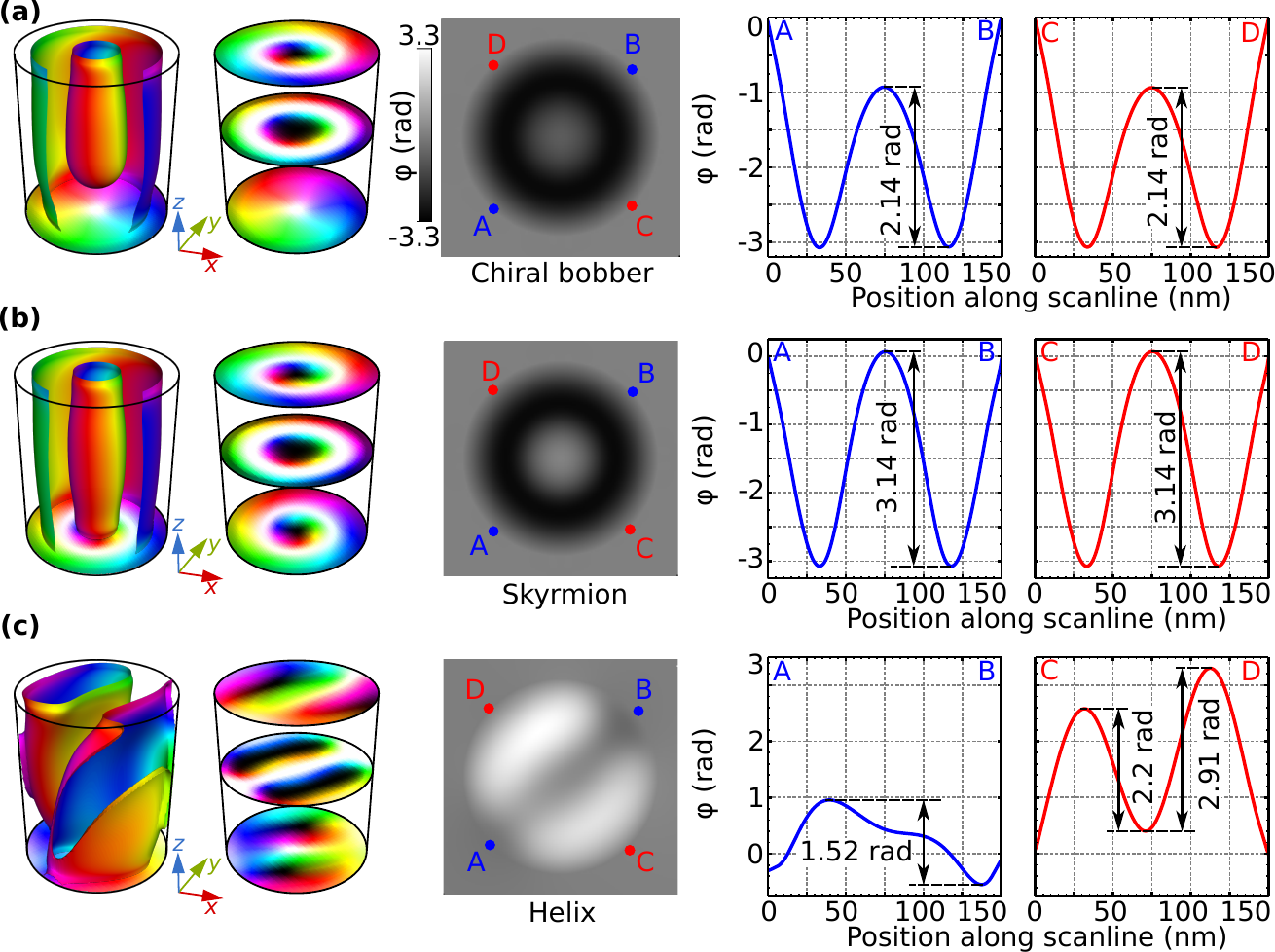}
\caption{\label{FigS1_Hologr_6} 
The calculated magnetic structures, phase shift images and line profiles for the chiral bobber (a), the skyrmion (b) and  the helix (c) in the FeGe nanocylinder with the soft magnetic surface layer of $6$ nm.}
\end{figure*}

\begin{figure*}
\includegraphics{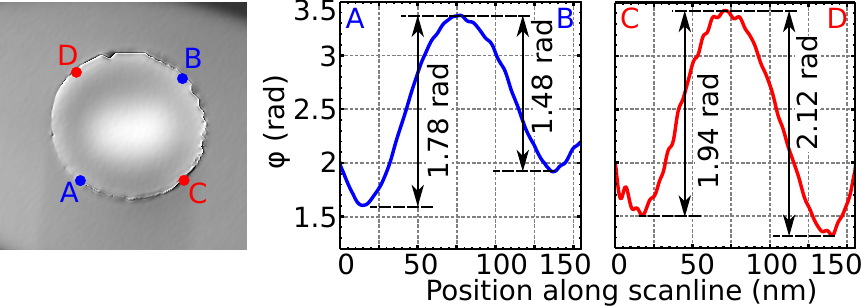}
\caption{\label{FigS2_Bobb} 
The experimental electron holography image of an FeGe nanocylinder, which is similar to the theoretical image for chiral bobber.}
\end{figure*}

 \begin{figure*}
\includegraphics{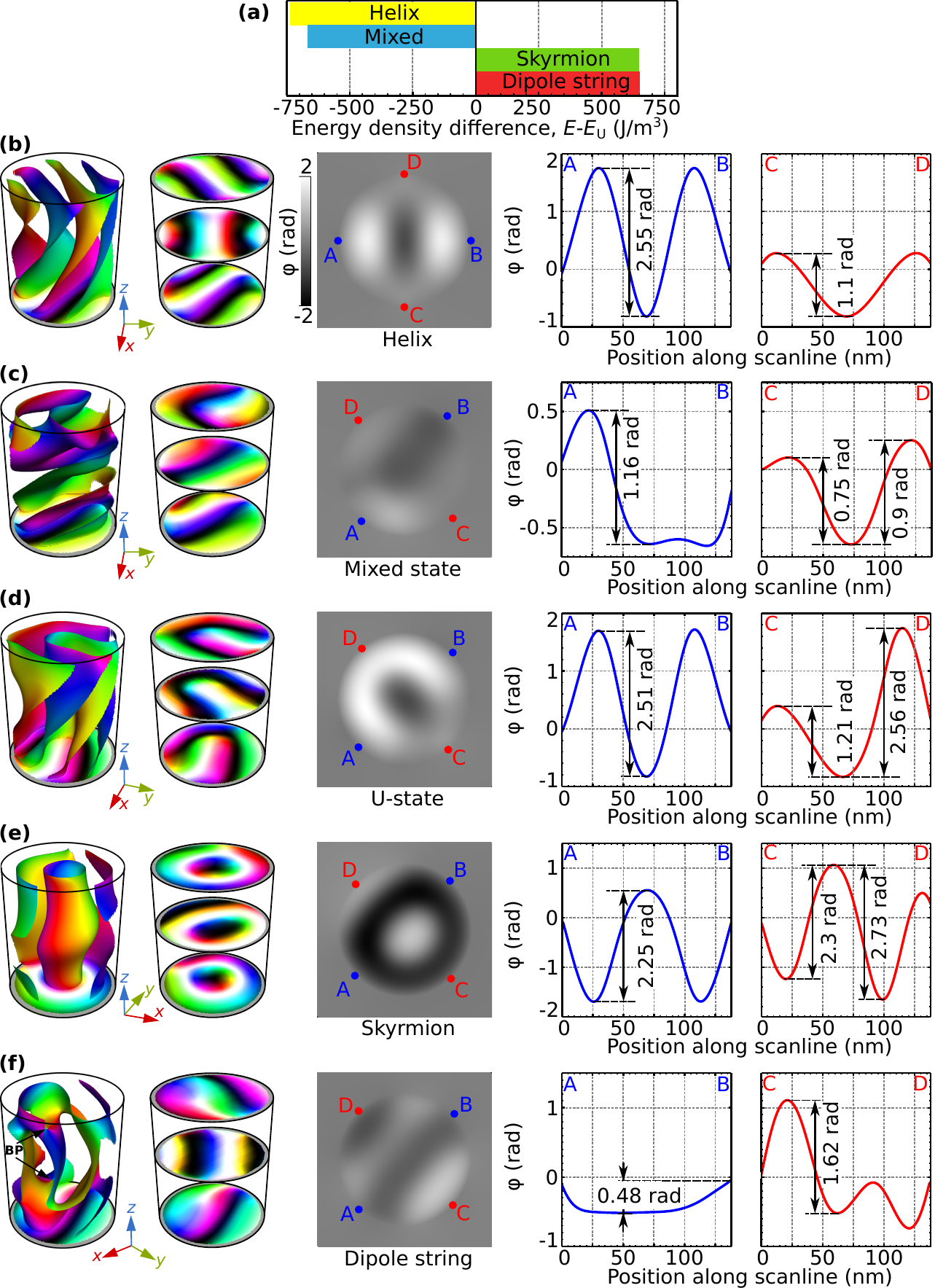}
\caption{\label{FigS4} 
The magnetic states, corresponding electron phase images and the energy densities for different states calculated for the FeGe nanocylinder of nominal size: 150 nm diameter and 180 nm height.
Here, wee assume that the surface damaged layer of thickness 6 nm is  nonmagnetic, meaning that the calculations were performed for the cylinder with actual size: 130 nm diameter and 150 nm height. (a) The energy density for different magnetic states depicted in (b)-(f) with respect to the energy of the U-state, $e_U=-39286.5$~Jm$^{-3}$.}
\end{figure*}